\documentclass[pra,twocolumn]{revtex4-1}

\usepackage{graphicx}% Include figure files
\usepackage{dcolumn}% Align table columns on decimal point
\usepackage{bm}% bold math
\usepackage{amssymb,amsmath,float}

\newcommand{\beq}{\begin{equation}}
\newcommand{\eeq}{\end{equation}}
\newcommand{\beqa}{\begin{eqnarray}}
\newcommand{\eeqa}{\end{eqnarray}}
\newcommand{\beqar}{\begin{eqnarray*}}
\newcommand{\eeqar}{\end{eqnarray*}}

\newcommand{\ket}[1]{|#1\rangle}

\newcommand{\ketbra}[2]{|#1\rangle\!\langle#2|}

\usepackage{theorem}

\def\squareforqed{\hbox{\rlap{$\sqcap$}$\sqcup$}}
\def\qed{\ifmmode\squareforqed\else{\unskip\nobreak\hfil
\penalty50\hskip1em\null\nobreak\hfil\squareforqed
\parfillskip=0pt\finalhyphendemerits=0\endgraf}\fi}
\def\endenv{\ifmmode\;\else{\unskip\nobreak\hfil
\penalty50\hskip1em\null\nobreak\hfil\;
\parfillskip=0pt\finalhyphendemerits=0\endgraf}\fi}

\begin{document}
\input epsf
\title{\bf \large When quantum games can be played classically: in support of van Enk-Pike's assertion}

\author{Berry Groisman} 
\address{ Department of Applied Mathematics and Theoretical Physics, University of Cambridge, Cambridge CB3 0WA, \textsc{United Kingdom}\\
and\\ Sidney Sussex College, Cambridge, CB2 3HU, \textsc{United Kingdom}}

\begin{abstract}
N.~Vyas and C.~Benjamin (arXiv:1701.08573[quant-ph]) propose a new mixed strategy for the (quantum) Hawk-Dove and Prisoners' Dilemma games and argue
that this strategy yields payoffs, which cannot be obtained in the corresponding classical games. They conclude that this refutes the earlier assertion by 
S.J.~van Enk and R.~Pike that the quantum equilibrium solution is present in a corresponding extended classical game. This paper argues that the scheme suggested by
N.~Vyas and C.~Benjamin changes the rules of the original game, and hence it does not refute the argument put forward by 
van Enk and Pike.
\end{abstract}

\maketitle

%\pacs{03.67-a, 03.65.Ud, 03.65.Ta}

\section{Introduction}
It is known that in many tasks quantum-mechanical protocols have genuine advantage over classical ones. However, there are cases where the advantage is only apparent. 
In the past, it has been proposed that extending classical games into the quantum domain can lead to emergence of novel features. In particular, in \cite{EWL_1999} it was argued that entanglement shared by the players in conjunction with a certain set of quantum operations gives rise to a new Nash equilibrium in the Prisoners' Dilemma (PD), thereby resolving the inherent contradiction of the classical game (as the new equilibrium is also Pareto-optimal). Their approach was criticised on the grounds that it does not solve the original classical game, because the quantum version of the game
is, in effect, equivalent to a different new classical game with extended payoff matrix \cite{vEP_2002}.   

Recently, Vyas and Benjamin \cite{VB_2017} claimed to have refuted van Enk-Pike's assertion. They used a mixed strategy on a maximally entangled state and argued that the solution yields payoffs which cannot be replicated in the classical game. They analyse the Hawk-Dove game in the main part of the paper and then draw an analogous conclusion in the Appendix. The aim of this comment is to show that there is a flaw in their argument, and hence that van Enk-Pike's assertion still stands. 

\section{The original protocol and van Enk-Pike's argument}\label{original protocol}
Here I briefly outline the scheme proposed in \cite{EWL_1999}. This scheme was subsequently modified \cite{LW_2014} for the Hawk-Dove game. 

At the first stage of the game, two qubits, denoted by $A$, $B$, are prepared by the arbitrator in a maximally entangled state 
\begin{equation}\label{in_PD}
 \ket{\psi_{in}^{PD}}=\frac{1}{\sqrt{2}}(\ket{00}-i\ket{11}.
 \end{equation}
 The state, which is known to Alice and Bob, is then distributed between the players: Alice and Bob get qubits $A$ and $B$ respectively.  

At the second stage, Alice and Bob both independently choose from a set of three operations, ${\cal O}\in\{\hat{C},\hat{D},\hat{Q}\}$, which correspond to the unitary operators $\mathbb{I}$, $iY$ and $iZ$ with $Y$ and $Z$ being the respective Pauli operators.  

At the third stage, the qubits are sent back to the arbitrator, who performs a disentangling unitary operation $\text{exp}\{-i\pi Y\otimes Y/4\}$, measures each qubit in
the computational basis and thus determines the payoffs. %The entire protocol is schematically presented in Fig. 1. Each bit is then measured in the computational basis. 

In the spirit of \cite{EWL_1999}, we assign the possible outcomes of the classical strategies (Cooperate or Defect of the original PD) to values $0$ and $1$ obtained as a result of the measurement, and the payoffs are calculated according to the payoff matrix \\
\begin{center}
\begin{tabular}{|c|cc}
\hline
$A\backslash B$&$0$&$1$\\
\hline
$0$&$(a,a)$&$(0,b)$\\[0.5ex]

$1$&$(b,0)$&$(c,c)$\\[0.5ex]

\end{tabular}
\end{center}
with $b>a>c>0$.

As it was pointed out in \cite{vEP_2002}, both players are choosing from three strategies, $C$, $D$ or $Q$, and the payoffs can be represented as being assigned according to the extended payoff matrix
\begin{center}
\begin{tabular}{c|ccc}

Alice$\backslash$Bob&$C$&$D$&$Q$\\
\hline
$C$&$(a,a)$&$~(0,b)~$&$(c,c)$\\[0.5ex]

$D$&$(b,0)$&$~(c,c)~$&$(0,b)$\\[0.5ex]

$Q$&$(c,c)$&$~(b,0)~$&$(a,a)$\\[0.5ex]

\end{tabular}
\end{center}
Hence, van Enk and Pike concluded that the quantum scheme of \cite{EWL_1999} does not solve the original classical game. Instead, it presents an extended game with new rules, which can be represented as a classical game with a new (extended) payoff matrix.

A similar argument can be made about the Hawk-Dove game.

\section{Vyas-Benjamin's argument}
Vyas and Benjamin \cite{VB_2017} introduce a mixed strategy on a maximally entangled state and argue that it yields payoffs which cannot be replicated in the classical Hawk-Dove game.
They extend their argument to the Prisoners' Dilemma. 

Again, I will refer to the Prisoners' Dilemma for the ease of reference. The core of my argument equally applies to the Hawk-Dove game, which takes the central stage in \cite{VB_2017}.
The random strategy is played using the initial state
\begin{equation}\label{in_R}
 \ket{\psi_{in}^{R}}=\frac{1}{\sqrt{2}}(\ket{00}+\ket{11}.
 \end{equation}
 Alice acts on $\ket{\psi_{in}^{R}}$ with identity operator $\mathbb{I}$ with probability $p$ and with flip operator $X$ with probability $1-p$. 
 Bob does the same with probabilities $(q,1-q)$. This results in $\ket{\psi_{f}^{R}}$.
 The payoffs are calculated as the mean values of payoff operators
\begin{equation}\label{PROJ}
\begin{split}
P_A&=3\ketbra{00}{00}+5\ketbra{10}{10}+\ketbra{11}{11}\\
P_B&=3\ketbra{00}{00}+5\ketbra{01}{01}+\ketbra{11}{11},
\end{split}
\end{equation}
 and hence 
 \begin{equation}\label{payoff}
\begin{split}
\$_A(p,q)&=Tr(P_A\ketbra{\psi_f^R}{\psi_f^R})\\
=\$_B(p,q)&=Tr(P_B\ketbra{\psi_f^R}{\psi_f^R})\\
&=\frac{1}{2}(4-2pq+p+q),
\end{split}
\end{equation}
which attains its maximal value of $\$(1,0)=\$(0,1)=2.5$. The authors then compare this result with the payoffs achieved in the original scheme (outlined in
Sec. \ref{original protocol}) with $a=3, b=5, c=1$ and 
refer to the (combined) quantum payoff matrix shown in Table \ref{VB_PD_PoM}.

\begin{table}[h!]
\begin{center}
\begin{tabular}{c|cccc}
Alice$\backslash$Bob&$C$&$D$&$Q$&$R$\\
\hline
$C$&$(3,3)$&$~(0,5)~$&$(1,1)$&$(2.5,2.5)$\\[0.5ex]

$D$&$(5,0)$&$~(1,1)~$&$(0,5)$&$(2.5,2.5)$\\[0.5ex]

$Q$&$(1,1)$&$~(5,0)~$&$(3,3)$&$(2.5,2.5)$\\[0.5ex]

$R$&$(2.5,2.5)$&$(2.5,2.5)$&$(2.5,2.5)$&$(2.5,2.5)$\\[0.5ex]
\end{tabular}
\caption{Combined payoff matrix constructed in \cite{VB_2017}.}.
\label{VB_PD_PoM}
\end{center}
\end{table}
This lead them to conclude that the payoffs corresponding to strategy $R$ cannot be replicated in the original classical game.

I can see two problems with their analysis. Firstly, Table \ref{VB_PD_PoM} does not seem to represent a payoff matrix of a single game. The entries of the $3\times 3$ submatrix 
correspond to the payoff achived when the parties follow the steps of Sec. \ref{original protocol}, which involves acting on the entangled state \eqref{in_PD}, which is different from
\eqref{in_R}, and using a particular disentangling operation before the payoffs are `measured' [through applications of \eqref{PROJ}]. In order to make a sensible comparison with
the fourth row and column of Table \ref{VB_PD_PoM}, we would need
to assume that Alice and Bob use the same resource (i.e. state \eqref{in_R}) and no disentangling operation is performed. But in this case the submatrix becomes\\
%\begin{table}
\begin{center}
\begin{tabular}{c|ccc}
Alice$\backslash$Bob&$C$&$D$&$Q$\\
\hline
$C$&$(2,2)$&$~(2.5,2.5)~$&$(2.5,2.5)$\\[0.5ex]

$D$&$(2.5,2.5)$&$~(2,2)~$&$(2.5,2.5)$\\[0.5ex]

$Q$&$(2.5,2.5)$&$~(2.5,2.5)~$&$(2,2)$\\[0.5ex]
\end{tabular}
%\caption{Combined payoff matrix constructed in \cite{VB_2017}.}.
%\label{VB_PD_PoM_updated}
\end{center}
%\end{table}
\noindent which replicates the payoffs achieved by strategy $R$.

Secondly, a special care should be taken of the fourth row and the fourth column of Table \ref{VB_PD_PoM}. Alice and Bob choose $p$ and $q$ to maximize their payoffs given by \eqref{payoff}. For the entry $R\backslash R$ the payoff of $(2.5,2.5)$ can only be guaranteed if both Alice and Bob agree on their choice of $(p,q)=(1,0)$ or $(0,1)$ 
in advance should they (individually) decide to play $R$. This choice will affect the rest of entries, so with $(p,q)=(0,1)$, say, the entry $C\backslash R$ will become $(2,2)$, because
Alice will act with $\hat{C}=\mathbb{I}$, which will not achieve the maximum of \eqref{payoff}.

To summarize, in order to work within the definition of a particular game, a mixed stratergy must operate within the space of the pure strategies of that game, which does not seem to be the 
case here. New strategies define new game.

\section{Conclusion}
By introducing the mixed strategy $R$, Vyas and Benjamin inevitibly change the rules of the game, hence they do not solve the original game. Moreover, their extended $4\times 4$ matrix
can be replicated in a classical game - the matrix defines the game. Thus, Vyas and Benjamin do not seem to have succeded in refuting the argument put forward by 
van Enk and Pike in \cite{vEP_2002}.

\vspace{0.5cm}
\begin{acknowledgments}
This work was generously supported by Sidney Sussex College, Cambridge. I am grateful to Colin Benjamin for his kind responses to my questions, and appreciate that he is 
likely to disagree with the conclusions of this paper.
\end{acknowledgments}

\end{document}